\documentclass[aps]{revtex4}

\usepackage{amsmath}

\def\ov#1{\overline{#1}}

\def\wt#1{\widetilde{#1}}
\def\vb#1{\mbox{\boldmath$#1$}}
\def\pd#1#2{\frac{\partial #1}{\partial #2}}
\def\fd#1#2{\frac{\delta #1}{\delta #2}}
\def\wh#1{\widehat{#1}}
\def\bdot{\,\vb{\cdot}\,}
\def\btimes{\,\vb{\times}\,}

\def\bhat{\wh{{\sf b}}}
\def\eq#1{\eqref{eq:#1}}
\def\exd{{\sf d}}

\newcommand{\bc}{\begin{center}}
\newcommand{\ec}{\end{center}}
\newcommand{\bt}{\begin{tabbing}}
\newcommand{\et}{\end{tabbing}} 
\newcommand{\be}{\begin{eqnarray*}}
\newcommand{\ee}{\end{eqnarray*}}

\begin{document}

\title{Guiding-center polarization and magnetization effects in gyrokinetic theory}

\author{Alain J.~Brizard}
\affiliation{Department of Physics, Saint Michael's College, Colchester, VT 05439, USA}

\begin{abstract}
Higher-order guiding-center polarization and magnetization effects are introduced in gyrokinetic theory by keeping first-order terms in background magnetic-field nonuniformity. These results confirm the consistency of the two-step perturbation analysis used in modern gyrokinetic theory.
\end{abstract}

\begin{flushright}
August 23, 2010
\end{flushright}

\maketitle

\section{Introduction}

Polarization and magnetization effects play a fundamental role in modern gyrokinetic theory \cite{Brizard_Hahm,Brizard_2010}. The standard form of modern gyrokinetic theory is derived by a two-step Lie-transform perturbation analysis that retains the effects of first-order ($\epsilon_{B}$) guiding-center drifts associated with the nonuniformity of the background magnetic field and gyrocenter effects up to second-order ($\epsilon_{\delta}^{2}$) in electromagnetic-field fluctuations that perturb the background guiding-center plasma. First-order $(\epsilon_{\delta})$ gyrocenter polarization and magnetization effects (which result from terms of order $\epsilon_{\delta}^{2}$ in the gyrocenter Hamiltonian) are fully retained in nonlinear gyrokinetic theory. Because the ordering parameters $\epsilon_{B}$ and $\epsilon_{\delta}$ are often comparable ($\epsilon_{B} \sim \epsilon_{\delta}$) in many practical applications of gyrokinetic theory, however, it is sometimes argued \cite{PC_2008} that first-order $(\epsilon_{B})$ guiding-center polarization and magnetization effects (which result from terms of order $\epsilon_{B}\,\epsilon_{\delta}$ in the gyrocenter Hamiltonian) should also be retained in nonlinear gyrokinetic theory for a consistent treatment of polarization and magnetization effects.

The two-step derivation of modern gyrokinetic theory is based on a sequence of two near-identity phase-space transformations ${\bf z}_{0} \rightarrow 
{\bf z} \rightarrow \ov{\bf z}$ from local particle coordinates ${\bf z}_{0} \equiv ({\bf x},v_{0\|},\mu_{0},\theta_{0})$ to guiding-center coordinates 
${\bf z} \equiv ({\bf X},v_{\|},\mu,\theta)$ and then to gyrocenter coordinates $\ov{{\bf z}} \equiv (\ov{\bf X},\ov{v}_{\|},\ov{\mu},\ov{\theta})$. The purpose of the guiding-center transformation \cite{RGL_83,Cary_Brizard} ${\bf z}_{0} \rightarrow {\bf z}$ (with small parameter $\epsilon_{B}$) is to asymptotically decouple the fast gyromotion of charged particles in a strong weakly-nonuniform $(\epsilon_{B} \ll 1)$ background magnetic field and construct the guiding-center magnetic moment $\mu \equiv \mu_{0} + \epsilon_{B}\,\mu_{1} + \cdots$ as an adiabatic invariant (from the local particle magnetic moment $\mu_{0} \equiv m|{\bf v}_{0\bot}|^{2}/2B$). The introduction of low-frequency electromagnetic-field fluctuations destroy the adiabatic invariance of the guiding-center magnetic moment $\mu$, which requires the gyrocenter transformation ${\bf z} \rightarrow \ov{\bf z}$ (with small parameter $\epsilon_{\delta}$) in order to restore the adiabatic invariance of the gyrocenter magnetic moment $\ov{\mu} \equiv \mu + \epsilon_{\delta}\,
\ov{\mu}_{1} + \cdots$.

Each transformation introduces a polarization charge density in the gyrokinetic Poisson equation and polarization and magnetization current densities in the gyrokinetic Amp\`{e}re equation \cite{Wang_Hahm}. These effects explicitly involve the generalized gyroradius $\ov{\vb{\rho}} \equiv {\bf x} - 
\ov{{\bf X}}$ defined as the displacement of the gyrocenter position $\ov{\bf X}$ from the particle position ${\bf x}$. As a result of the guiding-center and gyrocenter transformations, the generalized gyroradius
\begin{equation}
\ov{\vb{\rho}} \;\equiv\; \left( \vb{\rho}_{0{\rm gc}} \;+\; \epsilon_{B}\;\vb{\rho}_{1{\rm gc}} \;+\frac{}{} \cdots \right) \;+\;
\left( \epsilon_{\delta}\;\ov{\vb{\rho}}_{1{\rm gy}} \;+\frac{}{} \cdots \right)
\label{eq:rho_gc_gy}
\end{equation}
is decomposed into the guiding-center gyroradius $\vb{\rho}_{\rm gc} \equiv \vb{\rho}_{0{\rm gc}} + \epsilon_{B}\;\vb{\rho}_{1{\rm gc}}
+ \cdots$ and the gyrocenter gyroradius $\ov{\vb{\rho}}_{\rm gy} \equiv \epsilon_{\delta}\;\ov{\vb{\rho}}_{1{\rm gy}} + \cdots$ (which vanishes in the absence of fluctuations). As will be discussed below, polarization effects are associated with the gyrorangle-averaged displacement $\langle 
\ov{\vb{\rho}}\rangle$ while magnetization effects are associated with the gyrorangle-dependent displacement $\wt{\ov{\vb{\rho}}} \equiv \ov{\vb{\rho}} - \langle \ov{\vb{\rho}}\rangle$. 

The first-order $(\epsilon_{\delta})$ gyrocenter polarization and magnetization effects associated with $\ov{\vb{\rho}}_{1{\rm gy}}$ (which can also include terms of arbitrary order in $\epsilon_{B}$) have been discussed elsewhere \cite{Brizard_Hahm,Brizard_2010}. Moreover, we note that, since electromagnetic-field fluctuations satisfy the gyrokinetic ordering $|\vb{\rho}_{0{\rm gc}}\bdot\nabla| \sim 1$ (i.e., perpendicular wavelengths are of the same order as the lowest-order gyroradius), the guiding-center gyroradius-expansions considered in the present work apply only to the derivations of guiding-center polarization and magnetization effects. Hence, the first-order $(\epsilon_{\delta})$ gyrocenter polarization and magnetization effects will be ignored in what follows.

Because the lowest-order guiding-center gyroradius is explicitly gyroangle-dependent (i.e., $\langle\vb{\rho}_{0{\rm gc}}\rangle \equiv 0$), there is no zeroth-order term in the guiding-center polarization while the guiding-center magnetization has a non-vanishing zeroth-order term. First-order guiding-center polarization effects were investigated previously \cite{PM_85,PLS_85,Kaufman_86,Boghosian_87} outside the context of gyrokinetic theory. The purpose of the present paper is to investigate the higher-order polarization and magnetization effects introduced by the guiding-center gyroradius 
$\vb{\rho}_{\rm gc} = \vb{\rho}_{0{\rm gc}} + \epsilon_{B}\;\vb{\rho}_{1{\rm gc}} + \cdots$ within the standard format of nonlinear gyrokinetic theory. 

The remainder of the present paper is organized as follows. In Sec.~\ref{sec:gyro_pol_mag}, we introduce variational definitions of the reduced polarization and magnetization used in modern gyrokinetic theory based on functional derivatives of the gyrocenter Hamiltonian. In 
Sec.~\ref{sec:gc_pol_mag}, we explicitly make use of the higher-order correction $\vb{\rho}_{1{\rm gc}}$ to the guiding-center gyroradius 
$\vb{\rho}_{\rm gc}$ to derive explicit expressions for the first-order guiding-center polarization and magnetization. These results confirm that the two-step perturbation analysis used in modern gyrokinetic theory \cite{Brizard_Hahm} yield a consistent set of gyrokinetic Vlasov-Maxwell equations that include all first-order ($\epsilon_{B}$ and $\epsilon_{\delta}$) polarization and magnetization effects. 

\section{\label{sec:gyro_pol_mag}Gyrocenter Polarization and Magnetization}

The polarization effects in gyrokinetic theory are formally defined in terms of the relation (at a fixed position ${\bf r}$ at time $t$) between the particle charge density $\varrho({\bf r},t)$ and the gyrocenter charge density
$\varrho_{\rm gy}({\bf r},t)$:
\begin{equation}
\varrho \;\equiv\; \varrho_{\rm gy} \;-\; \nabla\bdot{\bf P}_{\rm gy},
\label{eq:pol_gy}
\end{equation}
where the polarization charge density $\varrho_{\rm pol} \equiv -\,\nabla\bdot{\bf P}_{\rm gy}$ serves as a definition of the gyrocenter polarization 
${\bf P}_{\rm gy}$. The gyrocenter charge density (summation over particle species is implied)
\begin{equation}
\varrho_{\rm gy} \;\equiv\; e\;\int\;\ov{F}\;d^{3}\ov{v}
\label{eq:gy_rho}
\end{equation}
is defined as a gyrocenter-velocity-space integral of the gyroangle-independent gyrocenter Vlasov distribution $\ov{F}$ (where $d^{3}\ov{v} \equiv 2\pi\,
\ov{\cal J}\,d\ov{v}_{\|}\,d\ov{\mu}$ with Jacobian $\ov{\cal J}$ to be defined below and the gyrocenter gyroangle integration has been performed explicitly). The gyrocenter polarization
\begin{equation}
{\bf P}_{\rm gy} \;\equiv\; {\bf P}_{\rm gc} \;+\; \epsilon_{\delta}\;{\bf P}_{1{\rm gy}} \;+\; \cdots,
\label{eq:gy_pol}
\end{equation}
on the other hand, is defined in terms of the guiding-center polarization
\begin{equation}
{\bf P}_{\rm gc} \;\equiv\; \epsilon_{B}\;{\bf P}_{1{\rm gc}} \;+\; \cdots,
\label{eq:gc_pol}
\end{equation}
which vanishes in a uniform magnetized plasma, and the first-order gyrocenter polarization $\epsilon_{\delta}{\bf P}_{1{\rm gy}}$, which vanishes in the absence of field fluctuations. 

The magnetization effects, on the other hand, are formally defined in terms of the relation (at a fixed position ${\bf r}$ at time $t$) between the particle current density ${\bf J}({\bf r},t)$ and the gyrocenter current density ${\bf J}_{\rm gy}({\bf r},t)$:
\begin{equation}
{\bf J} \;\equiv\; {\bf J}_{\rm gy} \;+\; \pd{{\bf P}_{\rm gy}}{t} \;+\; c\;\nabla\btimes{\bf M}_{\rm gy},
\label{eq:mag_gy}
\end{equation}
where the gyrocenter polarization current ${\bf J}_{\rm pol} \equiv \partial{\bf P}_{\rm gy}/\partial t$ is defined in terms of the gyrocenter polarization \eq{gy_pol} and the magnetization current ${\bf J}_{\rm mag} \equiv c\,\nabla\btimes{\bf M}_{\rm gy}$ serves as a definition of the gyrocenter magnetization ${\bf M}_{\rm gy}$. The gyrocenter current density
\begin{equation}
{\bf J}_{\rm gy} \;\equiv\; {\bf J}_{\rm gc} \;+\; \epsilon_{\delta}\;{\bf J}_{1{\rm gy}} \;+\; \cdots
\label{eq:gy_J}
\end{equation}
is defined in terms of the guiding-center current density
\begin{equation}
{\bf J}_{\rm gc} \;\equiv\; e\;\int\;\ov{F}\;\frac{d_{\rm gc}\ov{{\bf X}}}{dt}\;d^{3}\ov{v} \;=\; {\bf J}_{0{\rm gc}} \;+\; \epsilon_{B}\;
{\bf J}_{1{\rm gc}} \;+\; \cdots,
\label{eq:gc_J}
\end{equation}
which includes first-order guiding-center drifts, while the gyrocenter magnetization
\begin{equation}
{\bf M}_{\rm gy} \;\equiv\; {\bf M}_{\rm gc} \;+\; \epsilon_{\delta}\;{\bf M}_{1{\rm gy}} \;+\; \cdots
\label{eq:gy_mag}
\end{equation}
is defined in terms of the guiding-center magnetization
\begin{equation}
{\bf M}_{\rm gc} \;\equiv\; {\bf M}_{0{\rm gc}} \;+\; \epsilon_{B}\;{\bf M}_{1{\rm gc}} \;+\; \cdots,
\label{eq:gc_mag}
\end{equation}
where the non-vanishing zeroth-order guiding-center magnetization 
\begin{equation}
{\bf M}_{0{\rm gc}} \;\equiv\; -\;\left(\int\;\ov{\mu}\;\ov{F}\;d^{3}\ov{v} \right)\;\bhat
\label{eq:M0_gc}
\end{equation}
is a well-known result \cite{Kaufman_86,mag_0}.

Note that the definitions \eq{pol_gy} and \eq{mag_gy} ensure that the gyrokinetic Vlasov-Maxwell equations satisfy the gyrocenter charge conservation law $\partial\varrho_{\rm gy}/\partial t + \nabla\bdot{\bf J}_{\rm gy} = 0$, since polarization effects $\partial\varrho_{\rm pol}/\partial t + \nabla\bdot
{\bf J}_{\rm pol} \equiv 0$ conserve charge identically while the magnetization current density is divergenceless $\nabla\bdot{\bf J}_{\rm mag} \equiv 
0$. We also note that the quasineutrality condition $\varrho \equiv 0$ used in gyrokinetic theory \cite{Brizard_Hahm} can now be expressed
(up to order $\epsilon_{\delta}$ and $\epsilon_{B}$) as
\begin{equation}
\varrho_{\rm gy} \;-\; \nabla\bdot{\bf P}_{\rm gc} \;=\; \epsilon_{\delta}\;\nabla\bdot{\bf P}_{1{\rm gy}},
\label{eq:quasi_gy}
\end{equation}
where the left side contains the standard polarization effects associated with the lowest-order guiding-center gyroradius $\vb{\rho}_{0{\rm gc}}$ 
as well as the higher-order polarization effects associated with the first-order guiding-center gyroradius $\vb{\rho}_{1{\rm gc}}$ and gradients
$\nabla\vb{\rho}_{0{\rm gc}}$.

\subsection{Functional Definitions}

The gyrocenter polarization \eq{gy_pol} and magnetization \eq{gy_mag} can be defined as functionals of the gyrocenter Vlasov distribution $\ov{F}$ and functional derivatives of the gyrocenter Hamiltonian as follows \cite{Brizard_2000,Brizard_Hahm}. Here, the gyrocenter Hamiltonian 
\begin{equation}
H_{\rm gy} \;\equiv\; H_{\rm gc} \;+\; \epsilon_{\delta}\;H_{1{\rm gy}} \;+\; \epsilon_{\delta}^{2}\;H_{2{\rm gy}} \;+\; \cdots 
\label{eq:Hgy_def}
\end{equation}
is formally expressed as an asymptotic expansion in powers of $\epsilon_{B}$ and $\epsilon_{\delta}$. The unperturbed gyrocenter Hamiltonian is defined as the guiding-center Hamiltonian \cite{RGL_83}
\begin{equation}
H_{\rm gc} \;\equiv\; \frac{m}{2}\;\ov{v}_{\|}^{2} \;+\; \ov{\mu}\,B,
\label{eq:Hgc_def}
\end{equation}
where higher-order corrections ($\epsilon_{B}^{n}$, for $n \geq 1$) can be made to vanish \cite{Cary_Brizard}. The first-order gyrocenter Hamiltonian (in the Hamiltonian representation of gyrokinetic theory in which the guiding-center Poisson bracket is left unperturbed) is defined as 
\cite{Brizard_1989}
\begin{equation}
H_{1{\rm gy}} \;\equiv\; \left\langle e\;\Phi_{1}\left(\ov{{\bf X}} + \vb{\rho}_{\rm gc},t\right) \;-\; \frac{e}{c}\,{\bf A}_{1}\left(\ov{{\bf X}} + 
\vb{\rho}_{\rm gc},t\right)\bdot\left( \frac{d_{\rm gc}\ov{{\bf X}}}{dt} + \frac{d_{\rm gc}\vb{\rho}_{\rm gc}}{dt} \right) \right\rangle,
\label{eq:H1gy_def}
\end{equation}
where the first-order electromagnetic potentials $(\Phi_{1},{\bf A}_{1})$ represent the electromagnetic-field fluctuations that perturb the guiding-center plasma, and the guiding-center (unperturbed) evolution operator $d_{\rm gc}/dt \equiv \partial/\partial t + \{\;,\; H_{\rm gc}
\}_{\rm gc}$ is defined in terms of the guiding-center Hamiltonian \eq{Hgc_def} and the guiding-center Poisson bracket $\{\;,\;\}_{\rm gc}$ (to be defined below). The exact expression for the second-order gyrocenter Hamiltonian $H_{2{\rm gy}}$ has been given elsewhere 
\cite{Brizard_1989,Brizard_Hahm} and will not be needed in what follows since we will not be concerned with first-order $(\epsilon_{\delta})$ gyrocenter polarization and magnetization.

The first-order gyrocenter Hamiltonian \eq{H1gy_def}, which can be expanded as $H_{1{\rm gy}} \equiv H_{1{\rm gy}}^{(0)} + \epsilon_{B}\,
H_{1{\rm gy}}^{(1)} + \cdots$, contains the lowest-order terms \cite{Brizard_1989}
\begin{equation}
H_{1{\rm gy}}^{(0)} \;\equiv\; \left\langle e\;\Phi_{1{\rm gc}} \;-\; \frac{e}{c}\,{\bf A}_{1{\rm gc}}\bdot\left( \ov{v}_{\|}\,\bhat + \Omega\;\pd{\vb{\rho}_{0{\rm gc}}}{\ov{\theta}} \right) \right\rangle,
\label{eq:H1gy_0}
\end{equation}
where $\Phi_{1{\rm gc}} \equiv \Phi_{1}(\ov{\bf X} + \vb{\rho}_{0{\rm gc}},t)$ and ${\bf A}_{1{\rm gc}} \equiv {\bf A}_{1}(\ov{\bf X} + 
\vb{\rho}_{0{\rm gc}},t)$, as well as the first-order guiding-center corrections
\begin{eqnarray}
H_{1{\rm gy}}^{(1)} & \equiv & -\; \left\langle \vb{\rho}_{1{\rm gc}}\bdot\left[ e\;{\bf E}_{1{\rm gc}}
\;+\; \frac{e}{c} \left( \ov{v}_{\|}\,\bhat\btimes{\bf B}_{1\bot{\rm gc}} \;+\; \Omega\,\pd{\vb{\rho}_{0{\rm gc}}}{\ov{\theta}}\btimes
B_{1\|{\rm gc}}\,\bhat \right) \right]\right\rangle \nonumber \\
 &  &-\; \frac{e}{c}\,\left\langle{\bf A}_{1{\rm gc}}\bdot\left( \frac{d_{\rm gc}^{(1)}\ov{{\bf X}}}{dt} + 
\frac{d_{\rm gc}^{(1)}\vb{\rho}_{0{\rm gc}}}{dt} \right)\right\rangle 
\label{eq:H1gy_1}
\end{eqnarray}
where ${\bf E}_{1{\rm gc}} \equiv -\;\ov{\nabla}\Phi_{1{\rm gc}} - c^{-1}\partial{\bf A}_{1{\rm gc}}/\partial t$ and ${\bf B}_{1{\rm gc}} \equiv
\nabla\btimes{\bf A}_{1{\rm gc}} = B_{1\|{\rm gc}}\,\bhat + {\bf B}_{1\bot{\rm gc}}$, while $d_{\rm gc}^{(1)}\ov{\bf X}/dt$ and $d_{\rm gc}^{(1)}
\vb{\rho}_{0{\rm gc}}/dt$ denote first-order $(\epsilon_{B})$ corrections to $\ov{v}_{\|}\,\bhat$ and $\Omega\,\partial\vb{\rho}_{0{\rm gc}}/\partial\ov{\theta}$, respectively.

\subsubsection{Gyrocenter polarization}

The gyrocenter charge density and the gyrocenter polarization are defined as \cite{Brizard_2000}
\begin{equation}
\varrho_{\rm gy} \;-\; \nabla\bdot{\bf P}_{\rm gy} \;\equiv\; \epsilon_{\delta}^{-1}\;\int\;\ov{F}\;\fd{H_{\rm gy}}{\Phi_{1}({\bf r},t)}\;d^{6}\ov{z} 
\;=\; \int\;\ov{F}\;\fd{H_{1{\rm gy}}}{\Phi_{1}({\bf r},t)}\;d^{6}\ov{z} \;+\; \cdots,
\label{eq:rho_var_def}
\end{equation}
where guiding-center polarization effects \eq{gc_pol} are defined by the identity
\begin{equation}
-\; \nabla\bdot{\bf P}_{\rm gc} \;\equiv\; \int\;\ov{F}\;\left[\; e\frac{}{}\left\langle \delta^{3}(\ov{\bf X} + \vb{\rho}_{\rm gc}
- {\bf r})\right\rangle \;-\; e\;\delta^{3}(\ov{\bf X} - {\bf r}) \;\right]\;d^{6}\ov{z}.
\label{eq:rho_var_def_gc}
\end{equation}
By expanding Eq.~\eq{rho_var_def_gc} in powers of $\vb{\rho}_{\rm gc}$ and integrating by parts to eliminate $\delta^{3}(\ov{\bf X} - {\bf r})$, we obtain the guiding-center polarization
\begin{equation}
{\bf P}_{\rm gc} \;\equiv\; e\;\int\;\ov{F}\;\langle\vb{\rho}_{\rm gc}\rangle\;d^{3}\ov{v} \;-\; \nabla\bdot\left( \frac{e}{2}\; \int\;\ov{F}\;
\left\langle\vb{\rho}_{\rm gc}\frac{}{}\vb{\rho}_{\rm gc}\right\rangle\;d^{3}\ov{v} \right) \;+\; \cdots,
\label{eq:polgc_def}
\end{equation}
where dipole $(\langle\vb{\rho}_{\rm gc}\rangle$) and quadrupole $(\langle\vb{\rho}_{\rm gc}\vb{\rho}_{\rm gc}\rangle$) contributions are shown. By using the fact that $\langle\vb{\rho}_{0{\rm gc}}\rangle \equiv 0$, i.e., Eq.~\eq{polgc_def} has a vanishing zeroth-order term, the first-order guiding-center polarization is defined in terms of the functionals
\begin{equation}
{\bf P}_{1{\rm gc}} \;\equiv\; e\;\int \left\{\; \ov{F}\;\left[ \langle\vb{\rho}_{1{\rm gc}}\rangle \;-\; \nabla\bdot\left( \left\langle 
\frac{\vb{\rho}_{0{\rm gc}}\vb{\rho}_{0{\rm gc}}}{2}\right\rangle\right) \right] \;-\; \left\langle \frac{\vb{\rho}_{0{\rm gc}}
\vb{\rho}_{0{\rm gc}}}{2}\right\rangle\bdot\nabla \ov{F}\;\right\}d^{3}\ov{v},
\label{eq:P_1gc} 
\end{equation}
where we have grouped terms that directly involve the background magnetic-field nonuniformity and the term that directly involves the spatial gradient of the gyrocenter Vlasov distribution $\ov{F}$. The latter term (whose ordering is assumed to be comparable to $\epsilon_{B}$) is generated by the Taylor expansion of the integral
\begin{equation}
\int\;\ov{F}\;\left\langle \delta^{3}(\ov{{\bf X}} + \vb{\rho}_{0{\rm gc}} - {\bf r})\frac{}{}\right\rangle\;d^{6}\ov{z} \;=\;
\int\; \left( \ov{F} \;+\; \frac{\ov{\mu}B}{2\,m\Omega^{2}}\;\nabla_{\bot}^{2}\ov{F} \;+\; \cdots \right) \;d^{3}\ov{v},
\label{eq:gygc_density}
\end{equation}
where $\vb{\rho}_{0{\rm gc}}$ is assumed to be spatially uniform. We will return to Eq.~\eq{P_1gc} once we have obtained an expression for the first-order guiding-center gyroradius $\vb{\rho}_{1{\rm gc}}$ in the next Section.

\subsubsection{Gyrocenter magnetization}

The gyrocenter current density and the gyrocenter magnetization are defined as
\begin{equation}
{\bf J}_{\rm gy} \;+\; \pd{{\bf P}_{\rm gy}}{t} \;+\; c\;\nabla\btimes{\bf M}_{\rm gy} \;\equiv\; \epsilon_{\delta}^{-1}\;\int\;\ov{F}\; \left( -\,c\;
\fd{H_{\rm gy}}{{\bf A}_{1}({\bf r},t)}\right)\;d^{6}\ov{z} \;=\; \int\;\ov{F}\; \left( -\,c\;\fd{H_{1{\rm gy}}}{{\bf A}_{1}({\bf r},t)}\right)\;
d^{6}\ov{z} \;+\; \cdots,
\label{eq:J_var_def}
\end{equation}
where guiding-center magnetization effects \eq{gc_mag} are defined by the identity
\begin{equation}
{\bf J}_{\rm gc} \;+\; \pd{{\bf P}_{\rm gc}}{t} \;+\; c\;\nabla\btimes{\bf M}_{\rm gc} \;\equiv\; \int\;\ov{F}\;\left[\; e\,\left.\left.
\frac{d_{\rm gc}\ov{{\bf X}}}{dt}\;\right\langle \delta^{3}(\ov{\bf X} + \vb{\rho}_{\rm gc} - {\bf r})\right\rangle \;+\; e\;
\left\langle \frac{d_{\rm gc}\vb{\rho}_{\rm gc}}{dt}\;\delta^{3}(\ov{\bf X} + \vb{\rho}_{\rm gc} - {\bf r})\right\rangle \;\right]\;d^{6}\ov{z},
\label{eq:J_var_def_gc}
\end{equation}
with the guiding-center current density ${\bf J}_{\rm gc}$ defined by Eq.~\eq{gc_J}. After expanding Eq.~\eq{J_var_def_gc} in powers of 
$\vb{\rho}_{\rm gc}$ and carrying out several manipulations (see Appendix A for details), we obtain the guiding-center magnetization
\begin{equation}
{\bf M}_{\rm gc} \;\equiv\; \frac{e}{c}\;\int\; \ov{F} \left[ \frac{1}{2}\;\left\langle \vb{\rho}_{\rm gc}\btimes
\frac{d_{\rm gc}\vb{\rho}_{\rm gc}}{dt}\right\rangle \;+\; \langle\vb{\rho}_{\rm gc}\rangle\btimes\frac{d_{\rm gc}\ov{{\bf X}}}{dt}
\right]\;d^{3}\ov{v},
\label{eq:maggc_def}
\end{equation}
which is naturally split into an intrinsic contribution associated with $(1/2)\,\langle\vb{\rho}_{\rm gc}\btimes d_{\rm gc}\vb{\rho}_{\rm gc}/dt
\rangle$ and a moving-electric-dipole contribution $\langle\vb{\rho}_{\rm gc}\rangle\btimes d_{\rm gc}\ov{{\bf x}}/dt$. Note that the moving-magnetic-dipole contribution to the guiding-center polarization \eq{polgc_def} is a relativistic effect \cite{Brizard_JPCS} which falls outside the scope of the present work.

The guiding-center magnetization has a well-known zeroth-order intrinsic contribution \cite{mag_0}
\begin{equation}
{\bf M}_{0{\rm gc}} \;\equiv\; \frac{e}{2c}\;\int\; \ov{F} \left\langle \vb{\rho}_{0{\rm gc}}\btimes \left(\Omega\,
\pd{\vb{\rho}_{0{\rm gc}}}{\ov{\theta}}\right)\right\rangle\;d^{3}\ov{v},
\label{eq:M_0gc}
\end{equation}
from which we recover the classical result \eq{M0_gc}, where we used the lowest-order identities $\partial\vb{\rho}_{0{\rm gc}}/\partial\ov{\theta} 
\equiv \vb{\rho}_{0{\rm gc}}\btimes\bhat$ and $|\vb{\rho}_{0{\rm gc}}|^{2} \equiv 2\,\ov{\mu}B/(m\Omega^{2})$. The first-order contribution
\begin{eqnarray}
{\bf M}_{1{\rm gc}} & = & \frac{e}{c}\;\int\; \ov{F} \left[ \left\langle \wt{\vb{\rho}}_{1{\rm gc}}\btimes \left(\Omega\,
\pd{\vb{\rho}_{0{\rm gc}}}{\ov{\theta}}\right)\right\rangle \;+\; \frac{1}{2}\; \left\langle \vb{\rho}_{0{\rm gc}}\btimes 
\left( \frac{d_{\rm gc}^{(1)}\vb{\rho}_{0{\rm gc}}}{dt}\right)\right\rangle \right]\;d^{3}\ov{v} \nonumber \\
 &  &+\; \frac{e}{c}\;\int\; \ov{F}\; \left[ \langle\vb{\rho}_{1{\rm gc}}\rangle\btimes \left( \ov{v}_{\|}\;\bhat\right)\right]\;d^{3}\ov{v},
\label{eq:M1_gc}
\end{eqnarray}
on the other hand, is decomposed in terms of the first-order corrections (the first two terms) to the zeroth-order guiding-center intrinsic magnetization 
\eq{M_0gc}, while the third term yields the moving-electric-dipole contribution since it involves the gyroangle-averaged first-order guiding-center gyroradius $\langle\vb{\rho}_{1{\rm gc}}\rangle$ appearing in Eq.~\eq{P_1gc}.

\section{\label{sec:gc_pol_mag}Guiding-center Polarization and Magnetization}

In the present Section, we make use of the higher-order corrections to the guiding-center transformation \cite{RGL_83,Cary_Brizard} for the purpose of determining the first-order guiding-center gyroradius $\vb{\rho}_{1{\rm gc}}$ used in the first-order guiding-center polarization \eq{P_1gc} and the first-order guiding-center magnetization \eq{M1_gc}. We use Northrop's macroscopic interpretation \cite{Northrop} of the small parameter $\epsilon_{B}$ which, for finite macroscopic length $L_{B}$, allows us to use $e^{-1} \sim \epsilon$ as an ordering parameter. Hence, the guiding-center dynamical reduction is generated by the near-identity phase-space transformation
\begin{equation}
z^{\alpha} \;\equiv\; z_{0}^{\alpha} \;+\; \epsilon\;G_{1}^{\alpha} \;+\; \epsilon^{2} \left( G_{2}^{\alpha} \;+\; \frac{1}{2}\;G_{1}^{\beta}\;
\pd{G_{1}^{\alpha}}{z^{\beta}} \right) \;+\; \cdots,
\label{eq:zov_z}
\end{equation}
where $\epsilon \sim e^{-1}$ is used an ordering parameter and the phase-space vector field ${\sf G}_{n}$ is said to generate the phase-space transformation at order $n \geq 1$. 

\subsection{Guiding-center phase-space transformation}

The guiding-center coordinates $z^{\alpha} = ({\bf X}, v_{\|}, \mu, \theta)$ are defined (up to first order in $\epsilon_{B}$) as 
\cite{Cary_Brizard}
\begin{eqnarray}
{\bf X} & = & {\bf x} \;-\; \epsilon\;\vb{\rho}_{0} \;+\; \epsilon^{2}\;\left[ G_{2}^{{\bf x}} \;+\; \frac{1}{2}\;\vb{\rho}_{0}\bdot\nabla\vb{\rho}_{0} 
\;-\; \frac{1}{2} \left( G_{1}^{\mu}\;\pd{\vb{\rho}_{0}}{\mu_{0}} \;+\; G_{1}^{\theta}\;\pd{\vb{\rho}_{0}}{\theta_{0}} \right) \right], 
\label{eq:ovx_gc} \\
v_{\|} & = & v_{0\|} \left[ 1 \;-\; \epsilon\;\vb{\rho}_{0}\bdot\left(\bhat\bdot\nabla\bhat\right) \right] \;+\; \epsilon\;
\frac{\mu_{0} B}{m\Omega} \left( \tau \;+\; {\sf a}_{1}:\nabla\bhat \right), \label{eq:ovv_gc} \\
\mu & = & \mu_{0} \left[ 1 \;-\; \epsilon\;\rho_{0\|}\left( \tau \;+\; {\sf a}_{1}:\nabla\bhat \right) \;\right] \;+\; \epsilon\;
\vb{\rho}_{0}\bdot\left( \mu_{0}\;\nabla\ln B \;+\; \frac{m\,v_{0\|}^{2}}{B}\;\bhat\bdot\nabla\bhat \right), \label{eq:ovmu_gc} \\
\theta & = & \theta_{0} \;-\; \epsilon\;\vb{\rho}_{0}\bdot{\bf R} \;+\; \epsilon\;\rho_{0\|}\;\left( {\sf a}_{2}:\nabla\bhat \right) \;+\; 
\epsilon\;\pd{\vb{\rho}_{0}}{\theta_{0}}\bdot\left( \nabla\ln B \;+\; \frac{mv_{0\|}^{2}}{2\,\mu_{0} B}\;\bhat\bdot\nabla\bhat\right), 
\label{eq:ovtheta_gc}
\end{eqnarray}
where $\rho_{0\|} \equiv v_{0\|}/(m\Omega)$ denotes the ``parallel'' gyroradius, $\tau \equiv \bhat\bdot\nabla\btimes\bhat$ denotes the background magnetic torsion, ${\bf R}$ denotes the gyrogauge vector field, and expressions for the gyroangle-dependent dyadic tensors $({\sf a}_{1}, {\sf a}_{2})$ are not needed in what follows. Here, the second-order spatial component $G_{2}^{{\bf x}}$ is expressed as
\begin{equation}
G_{2}^{{\bf x}} \;=\; G_{2\|}\;\bhat \;+\; \rho_{\|}\,\tau \;\vb{\rho}_{0} \;+\; \frac{1}{2} \left( G_{1}^{\mu} \;-\frac{}{} 
\mu\;\vb{\rho}_{0}\bdot\nabla\ln B \right) \pd{\vb{\rho}_{0}}{\mu_{0}} \;+\; \frac{1}{2} \left( G_{1}^{\theta} \;+\frac{}{} \vb{\rho}_{0}\bdot{\bf R} \right) \pd{\vb{\rho}_{0}}{\theta_{0}}.
\label{eq:G2_x}
\end{equation}
where
\begin{equation}
G_{2\|} \;\equiv\; \bhat\bdot G_{2}^{{\bf x}} \;=\; 2\;\rho_{0\|}\;\pd{\vb{\rho}_{0}}{\theta}\bdot\left(\bhat\bdot\nabla\bhat
\right) \;+\; \frac{\mu_{0}\,B}{m\Omega^{2}}\;\left({\sf a}_{2}:\nabla\bhat\right).
\label{eq:G2_par}
\end{equation}
The first-order components $(G_{1}^{\mu}, G_{1}^{\theta})$, defined in Eqs.~\eq{ovmu_gc}-\eq{ovtheta_gc} and used in Eqs.~\eq{ovx_gc} and \eq{G2_x}, are
\begin{eqnarray}
G_{1}^{\mu} & \equiv & \vb{\rho}_{0}\bdot\left( \mu_{0}\;\nabla\ln B \;+\; \frac{m\,v_{0\|}^{2}}{B}\;\bhat\bdot\nabla\bhat \right) \;-\; \mu_{0}\; 
\rho_{0\|}\; \left( \tau \;+\; {\sf a}_{1}:\nabla\bhat \right),
\label{eq:G1_mu} \\
G_{1}^{\theta} & \equiv & -\;\vb{\rho}_{0}\bdot{\bf R} \;+\; \rho_{0\|}\;\left( {\sf a}_{2}:\nabla\bhat \right) \;+\; \pd{\vb{\rho}_{0}}{\theta_{0}}
\bdot\left( \nabla\ln B \;+\; \frac{m\,v_{0\|}^{2}}{2\,\mu_{0} B}\;\bhat\bdot\nabla\bhat\right).
\label{eq:G1_theta}
\end{eqnarray}
The remaining first-order component defined in Eq.~\eq{ovv_gc}
\begin{equation}
G_{1}^{v_{\|}} \;\equiv\; -\;v_{0\|}\;\vb{\rho}_{0}\bdot\left(\bhat\bdot\nabla\bhat\right) \;+\; \frac{\mu_{0} B}{m\Omega} \left( \tau \;+\; 
{\sf a}_{1}:\nabla\bhat \right)
\label{eq:G1_vpar}
\end{equation}
ensures that the first-order correction 
\begin{equation}
H_{1{\rm gc}} \;\equiv\; -\;G_{1}^{\alpha}\,\pd{H_{\rm gc}}{z^{\alpha}} \;=\; \mu\;\vb{\rho}_{0}\bdot\nabla B \;-\; mv_{\|}\;G_{1}^{v_{\|}} \;-\;
B\;G_{1}^{\mu} \;\equiv\; 0
\label{eq:H1_gc}
\end{equation}
to the guiding-center Hamiltonian \eq{Hgc_def} vanishes identically. The same construction algorithm can be applied at higher order (i.e., 
$H_{n{\rm gc}} \equiv -\,G_{n}^{\alpha}\partial H_{\rm gc}/\partial z^{\alpha} \equiv 0$ for $n \geq 1$), so that the simplicity of the guiding-center Hamiltonian can be preserved in the form of Eq.~\eq{Hgc_def} to all orders in $\epsilon$ \cite{RGL_83}.

The Jacobian for the guiding-center phase-space transformation \eq{ovx_gc}-\eq{ovtheta_gc} is constructed from the Jacobian for the local particle phase-space coordinates
${\cal J}_{0} = B/m$ according to the formula
\begin{eqnarray} 
{\cal J} & \equiv & {\cal J}_{0} \;-\; \epsilon\;\pd{}{z^{\alpha}} \left( {\cal J}_{0}\;G_{1}^{\alpha} \right) \;+\; \cdots \;=\; \frac{B}{m} \;+\; \epsilon \left[\; \nabla\bdot\left( \frac{B}{m}\;\vb{\rho}_{0}\right) \;-\; \frac{B}{m} \left( \pd{G_{1}^{v_{\|}}}{v_{\|}} \;+\; \pd{G_{1}^{\mu}}{\mu} 
\;+\; \pd{G_{1}^{\theta}}{\theta} \right) \;\right] \;+\; \cdots \nonumber \\
 & = & \frac{B}{m} \left( 1 \;+\; \epsilon\;\rho_{\|}\;\tau \;+\; \cdots \right) \;\equiv\; \frac{B_{\|}^{*}}{m}.
\label{eq:Jac_gc}
\end{eqnarray}
Hence,the near-identity guiding-center phase-space transformation is noncanonical since $B_{\|}^{*} \neq B$. 

As a result of the guiding-center transformation \eq{ovx_gc}-\eq{ovtheta_gc}, the guiding-center phase-space Lagrangian is expressed as
\begin{equation}
\Gamma_{\rm gc} \;\equiv\; \left( \frac{e}{\epsilon\,c}\,{\bf A} \;+\; mv_{\|}\;\bhat \;-\; \epsilon\;\mu\,\frac{B}{\Omega}\;{\bf R}^{*}\right)
\bdot\exd{\bf X} \;+\; \epsilon\;\mu\,\frac{B}{\Omega}\;\exd\theta \;-\; H_{\rm gc}\;\exd t,
\label{eq:PSL_gc}
\end{equation}
where ${\bf R}^{*} \equiv {\bf R} + (\tau/2)\,\bhat$ and the guiding-center Hamiltonian $H_{\rm gc}$ is given by Eq.~\eq{Hgc_def}. The guiding-center Poisson bracket $\{\;,\;\}_{\rm gc}$ constructed from the symplectic part of the guiding-center phase-space Lagrangian \eq{PSL_gc} is \cite{Cary_Brizard}
\begin{equation}
\{ F,\; G\}_{\rm gc} \;=\; \epsilon^{-1}\;\frac{\Omega}{B} \left(\pd{F}{\theta}\,\pd{G}{\mu} - \pd{F}{\mu}\,\pd{G}{\theta} \right) \;+\; 
\frac{{\bf B}^{*}}{m\,B_{\|}^{*}}\bdot\left( \nabla^{*} F\,\pd{G}{v_{\|}} - \pd{F}{v_{\|}}\,\nabla^{*} G \right) \;-\; \epsilon\;
\frac{c\bhat}{e\,B_{\|}^{*}}\bdot\nabla^{*}F\btimes\nabla^{*}G,
\label{eq:gc_PB}
\end{equation}
where $\nabla^{*} \equiv \nabla + {\bf R}^{*}\,\partial/\partial\theta$, 
\begin{equation}
{\bf B}^{*} \;\equiv\; {\bf B} \;+\; B \left( \epsilon\;\rho_{\|}\,\nabla\btimes\bhat \;-\; \epsilon^{2}\;\frac{\mu\,B}{m\,\Omega^{2}}\,
\nabla\btimes{\bf R}^{*} \;+\; \cdots \right),
\label{eq:Bstar_def}
\end{equation}
and $B_{\|}^{*} \equiv \bhat\bdot{\bf B}^{*}$. Note that, under the gyrogauge transformation $\theta \rightarrow \theta^{\prime} \equiv \theta + 
\psi({\bf X})$, the vector ${\bf R}$ is gyrogauge-dependent (i.e., ${\bf R} \rightarrow {\bf R}^{\prime} \equiv {\bf R} + \nabla\psi$), while the curl of ${\bf R}$ is gyrogauge-invariant (i.e., $\nabla\btimes{\bf R}^{\prime} = \nabla\btimes{\bf R}$). Hence, the guiding-center phase-space Lagrangian 
\eq{PSL_gc} and the guiding-center Poisson bracket \eq{gc_PB} are both gyrogauge-invariant \cite{RGL_83} since the combinations $\exd\theta - {\bf R}\bdot\exd{\bf X}$ and $\nabla + {\bf R}\,\partial/\partial\theta$ are gyrogauge-invariant.

Lastly, we can now write expressions for the velocities $d_{\rm gc}{\bf X}/dt$ and $d_{\rm gc}\vb{\rho}_{0}/dt$ to be used in evaluating the first-order guiding-center magnetization \eq{maggc_def}. First, the guiding-center velocity
\begin{equation}
\frac{d_{\rm gc}{\bf X}}{dt} \;\equiv\; \{ {\bf X},\; H_{\rm gc}\}_{\rm gc} \;=\; v_{\|}\,\bhat \;+\; \frac{c\bhat}{eB_{\|}^{*}}\btimes\left(\mu\;
\nabla B \;+\frac{}{} mv_{\|}^{2}\;(\bhat\bdot\nabla\bhat)\right) \;\equiv\; v_{\|}\,\bhat \;+\; \epsilon_{B}\;\frac{d_{\rm gc}^{(1)}{\bf X}}{dt}
\label{eq:dgc_x}
\end{equation}
is expressed in terms of the zeroth-order motion along a magnetic-field line and first-order guiding-center drifts. Second, the gyration velocity
\begin{equation}
\frac{d_{\rm gc}\vb{\rho}_{0}}{dt} \;\equiv\; \{ \vb{\rho}_{0},\; H_{\rm gc}\}_{\rm gc} \;=\; \Omega\;\pd{\vb{\rho}_{0}}{\theta} \;+\; 
v_{\|}\,\bhat\bdot\nabla^{*}\vb{\rho}_{0} \;+\; \cdots \;\equiv\; \Omega\;\pd{\vb{\rho}_{0}}{\theta} \;+\; \epsilon_{B}\;
\frac{d_{\rm gc}^{(1)}\vb{\rho}_{0}}{dt} \;+\; \cdots
\label{eq:dgc_rho}
\end{equation}
is expressed in terms of the zeroth-order perpendicular particle velocity and its first-order correction. The guiding-center velocities \eq{dgc_x} and
\eq{dgc_rho} appear in the first-order gyrocenter Hamiltonian \eq{H1gy_def} and, consequently, they contribute to the guiding-center magnetization 
\eq{maggc_def}.

\subsection{First-order corrections to the guiding-center gyroradius}

So far we have not made a distinction between the gyroradius in particle phase space (labeled $\vb{\rho}$) and the gyroradius in guiding-center phase space (labeled $\vb{\rho}_{\rm gc}$), which are respectively defined as
\begin{equation}
\left. \begin{array}{rcl}
\vb{\rho} & \equiv & {\bf x} \;-\; {\sf T}_{\rm gc}{\bf X} \;=\; \vb{\rho}_{0} \;+\; \epsilon\;\vb{\rho}_{1} \;+\; \cdots \\
 &  & \\
\vb{\rho}_{\rm gc} & \equiv & {\sf T}_{\rm gc}^{-1}{\bf x} \;-\; {\bf X} \;=\; \vb{\rho}_{0{\rm gc}} \;+\; \epsilon\;\vb{\rho}_{1{\rm gc}} \;+\; \cdots
\end{array} \right\}.
\label{eq:rho_rhobar_def}
\end{equation}
In Eq.~\eq{rho_rhobar_def}, the pull-back operator ${\sf T}_{\rm gc}$ transforms a function on guiding-center phase space into a function on particle phase space, while the push-forward operator ${\sf T}_{\rm gc}^{-1}$ transforms a function on particle phase space into a function on guiding-center phase space. From these definitions, we therefore obtain the relation between the particle and guiding-center gyroradii
\begin{equation}
\vb{\rho}_{\rm gc} \;\equiv\; {\sf T}_{\rm gc}^{-1}\vb{\rho} \;=\; \vb{\rho} \;-\; \epsilon\;G_{1}^{\alpha}\,\pd{\vb{\rho}}{z^{\alpha}} \;+\; 
\cdots \;\equiv\; \vb{\rho}_{0{\rm gc}} \;+\; \epsilon\;\vb{\rho}_{1{\rm gc}} \;+\; \cdots,
\label{eq:rho_rhobar}
\end{equation}
so that the lowest-order guiding-center gyroradius is $\vb{\rho}_{0{\rm gc}} \equiv \vb{\rho}_{0}$ is identical in both phase spaces. The first-order gyroradii $\vb{\rho}_{1}$ and $\vb{\rho}_{1{\rm gc}} \equiv \vb{\rho}_{1} - G_{1}^{\alpha}\,\partial\vb{\rho}_{0}/\partial z^{\alpha}$ are not identical, however, and it is important to use the proper first-order gyroradius in order to obtain the correct first-order polarization and magnetization.

In particle phase space, the guiding-center position ${\bf X}$ is expressed in terms of the guiding-center transformation (\ref{eq:ovx_gc}) as ${\bf X} = {\bf x} - \epsilon\,\vb{\rho}_{0} - \epsilon^{2}\vb{\rho}_{1} + \cdots \equiv {\sf T}_{\rm gc}{\bf x}$, where the first-order displacement 
$G_{1}^{{\bf x}} \equiv -\,\vb{\rho}_{0}$ defines the lowest-order gyroradius, while the second-order correction is
\begin{eqnarray}
\vb{\rho}_{1} & = & -\;G_{2}^{{\bf x}} \;-\; \frac{1}{2} \left[\; \vb{\rho}_{0}\bdot\nabla\vb{\rho}_{0} \;-\; \left( G_{1}^{\mu}\;
\pd{\vb{\rho}_{0}}{\mu} \;+\; G_{1}^{\theta}\;\pd{\vb{\rho}_{0}}{\theta} \right) \;\right] \nonumber \\
 & \equiv &  \left(\frac{1}{2}\,\vb{\rho}_{0}\bdot\nabla\ln B \;-\; \rho_{\|}\,\tau\right)\;\vb{\rho}_{0} \;+\; \left(\frac{1}{2}\,\vb{\rho}_{0}\bdot\nabla\bhat\bdot\vb{\rho}_{0} \;-\; G_{2\|} \right)\;\bhat.
\label{eq:rho1_particle} 
\end{eqnarray}
The gyroangle-averaged particle gyroradius \eq{rho1_particle} is
\begin{equation} 
\langle\vb{\rho}_{1}\rangle \;=\; \frac{\mu B}{2\,m\Omega^{2}}\;\left[ \left(\nabla\bdot\bhat\right)\;\bhat \;+\; \nabla_{\bot}\ln B \right] \;\equiv\; 
-\;\nabla\bdot\left( \left\langle \frac{\vb{\rho}_{0}\vb{\rho}_{0}}{2}\right\rangle \right) \;-\; \frac{\mu B\,(\bhat\bdot\nabla\bhat)}{2\;m
\,\Omega^{2}},
\label{eq:rho_av}
\end{equation}
where we used $\langle G_{2\|}\rangle \equiv 0$ from Eq.~\eq{G2_par}.

In guiding-center phase space, on the other hand, where the particle position ${\bf x}$ is expressed in terms of the inverse guiding-center transformation as ${\bf x} = {\bf X} + \vb{\rho}_{0{\rm gc}} + \vb{\rho}_{1{\rm gc}} + \cdots \equiv {\sf T}_{\rm gc}^{-1}{\bf X}$, the first-order correction is
\begin{eqnarray}
\vb{\rho}_{1{\rm gc}} & = & -\,G_{2}^{{\bf x}} \;+\; \frac{1}{2} \left[\; \vb{\rho}_{0}\bdot\nabla\vb{\rho}_{0} \;-\; \left( G_{1}^{\mu}\;
\pd{\vb{\rho}_{0}}{\mu} \;+\; G_{1}^{\theta}\;\pd{\vb{\rho}_{0}}{\theta} \right) \;\right] \nonumber \\
 & \equiv & -\;\left( G_{1}^{\theta} \;+\frac{}{} \vb{\rho}_{0}\bdot{\bf R}\right) \pd{\vb{\rho}_{0}}{\theta} \;-\; \left( G_{1}^{\mu} \;+\frac{}{} 
2\,\mu\,\rho_{\|}\;\tau\right) \;\pd{\vb{\rho}_{0}}{\mu} \;-\; \left( G_{2\|} \;+\; \frac{1}{2}\;\vb{\rho}_{0}\bdot\nabla\bhat\bdot\vb{\rho}_{0},
\right)\;\bhat
\label{eq:rho1_gc} 
\end{eqnarray}
where the same generating vector fields $({\sf G}_{1}, {\sf G}_{2}, \cdots)$ are used and we inserted the identity
\[ \vb{\rho}_{0}\bdot\nabla\vb{\rho}_{0} \;\equiv\; -\;\left(\mu\,\vb{\rho}_{0}\bdot\nabla\ln B\right)\;\pd{\vb{\rho}_{0}}{\mu} \;-\; \left( \vb{\rho}_{0}
\bdot{\bf R}\right)\;\pd{\vb{\rho}_{0}}{\theta} \;-\; \left(\vb{\rho}_{0}\bdot\nabla\bhat\bdot\vb{\rho}_{0}\right)\;\bhat \]
in obtaining the last expression for $\vb{\rho}_{1{\rm gc}}$.

Lastly, we note that the first-order particle and guiding-center gyroradii \eq{rho1_particle} and \eq{rho1_gc} are both gyrogauge-invariant. Moreover, according to Eqs.~\eq{P_1gc} and \eq{M1_gc}, it is the guiding-center first-order gyroradius vector \eq{rho1_gc} that must be used and, hence, we will use $\langle\vb{\rho}_{1{\rm gc}}\rangle$ to compute the first-order guiding-center polarization \eq{P_1gc} and $\wt{\vb{\rho}}_{1{\rm gc}} \equiv 
\vb{\rho}_{1{\rm gc}} - \langle\vb{\rho}_{1{\rm gc}}\rangle$ to compute the first-order guiding-center magnetization \eq{M1_gc}.

\subsection{First-order Guiding-center Polarization}

The gyroangle-averaged guiding-center gyroradius \eq{rho1_gc} is
\begin{equation}
\langle\vb{\rho}_{1{\rm gc}}\rangle \;\equiv\; \frac{\bhat}{\Omega}\btimes\frac{d_{\rm gc}{\bf X}}{dt} \;+\; \nabla\bdot\left( \left\langle 
\frac{\vb{\rho}_{0}\vb{\rho}_{0}}{2}\right\rangle \right) \;+\; \frac{\mu B\,(\bhat\bdot\nabla\bhat)}{2\;m\,\Omega^{2}},
\label{eq:ovrho1_ave}
\end{equation}
where we used Eq.~\eq{rho_av}. When we combine these results into the first-order guiding-center polarization \eq{P_1gc}, we obtain
\begin{equation}
{\bf P}_{1{\rm gc}} \;=\; \frac{\bhat}{\Omega}\btimes\left[\; \int\;\ov{F}\;\left( e\;\frac{d_{\rm gc}\ov{{\bf X}}}{dt}\right)\;d^{3}\ov{v} \;+\; c\;\nabla\btimes\left(\frac{1}{2}\;\int\;\left(-\,\ov{\mu}\,\bhat\right)\;\ov{F}\;d^{3}\ov{v} \right) \;\right].
\label{eq:P1gc_def}
\end{equation}
This expression combines the classical first-order (dipole) guiding-center polarization obtained previously \cite{Kaufman_86} as well as the first-order quadrupole guiding-center polarization.

\subsection{First-order Guiding-center Magnetization}

The calculation of the first-order guiding-center magnetization \eq{M1_gc} requires the guiding-center push-forward of the particle velocity ${\bf v} = 
d{\bf x}/dt$, which is defined as
\begin{equation}
{\bf V}_{\rm gc} \;\equiv\; {\sf T}_{\rm gc}^{-1}{\bf v} \;\equiv\; \frac{d_{\rm gc}{\bf X}}{dt} \;+\; \frac{d_{\rm gc}\vb{\rho}_{\rm gc}}{dt} \;=\; 
{\bf V}_{0{\rm gc}} \;+\; \epsilon\;{\bf V}_{1{\rm gc}} \;+\; \cdots,
\end{equation}
where 
\[ {\bf V}_{0{\rm gc}} \;\equiv\; \frac{d_{\rm gc}^{(0)}{\bf X}}{dt} \;+\; \frac{d_{\rm gc}^{(0)}\vb{\rho}_{0{\rm gc}}}{dt} \;=\; v_{\|}\,\bhat 
\;+\; \Omega\,\pd{\vb{\rho}_{0{\rm gc}}}{\theta} \]
denotes the lowest-order guiding-center (particle) velocity and its first-order correction is
\begin{equation}
{\bf V}_{1{\rm gc}} \;\equiv\; \langle{\bf V}_{1{\rm gc}}\rangle \;+\; \wt{{\bf V}}_{1{\rm gc}},
\label{eq:v1_def}
\end{equation}
where the gyroangle-independent part is
\begin{equation}
\langle{\bf V}_{1{\rm gc}}\rangle \;\equiv\; \frac{d_{\rm gc}^{(1)}{\bf X}}{dt} \;=\; \left( \bhat\btimes\frac{d_{\rm gc}{\bf X}}{dt}\right)\btimes\bhat
\label{eq:v1_av}
\end{equation}
and the gyroangle-dependent part is
\begin{equation}
\wt{{\bf V}}_{1{\rm gc}} \;\equiv\; \frac{d_{\rm gc}^{(0)}\wt{\vb{\rho}}_{1{\rm gc}}}{dt} \;+\; \frac{d_{\rm gc}^{(1)}\vb{\rho}_{0{\rm gc}}}{dt} \;=\; 
\Omega\;\left( \pd{\wt{\vb{\rho}}_{1{\rm gc}}}{\theta} \;+\; \rho_{\|}\bhat\bdot\nabla^{*}\vb{\rho}_{0{\rm gc}} \right).
\label{eq:v1_tilde}
\end{equation}
Here, we easily show that the guiding-center velocity ${\bf V}_{\rm gc}$ satisfies the following identities: $\langle{\bf V}_{\rm gc}\rangle \equiv 
d_{\rm gc}{\bf X}/dt$ and $H_{\rm gc} = (m/2)\,|{\bf V}_{\rm gc}|^{2}$. 

Lastly, by inserting Eqs.~\eq{rho1_gc}-\eq{ovrho1_ave} and \eq{v1_tilde} into Eq.~\eq{M1_gc}, we obtain the first-order guiding-center magnetization
\begin{equation}
{\bf M}_{1{\rm gc}} \;=\; \int\;\ov{\rho}_{\|}\,\ov{F} \left[\; \ov{\mu} \left( \tau\;\bhat \;+\; \frac{3}{2}\,B\;\nabla\btimes\left(B^{-1}\,\bhat
\right) \right) \;+\; \frac{m\,\ov{v}_{\|}^{2}}{B}\;\bhat\btimes\left(\bhat\bdot\nabla\bhat\right) \;\right]\;d^{3}\ov{v}.
\label{eq:M1gc_def}
\end{equation}
We note, however, that the first-order guiding-center magnetization \eq{M1gc_def} vanishes if the gyrocenter Vlasov distribution $\ov{F}$ is a Maxwellian in $\ov{v}_{\|}$. In most applications of gyrokinetic theory, the zeroth-order guiding-center magnetization \eq{M_0gc} is therefore sufficient.

\section{\label{sec:sum}Summary}

We have presented the derivation of guiding-center polarization and magnetization effects as a simple extension of the standard form of modern gyrokinetic theory by keeping higher-order corrections to the guiding-center gyroradius \eq{rho1_gc} in the first-order gyrocenter Hamiltonian
\eq{H1gy_def} [e.g., Eq.~\eq{H1gy_1}]. 

From the variational derivations of the guiding-center polarization and magnetization, we have recovered the classical first-order guiding-center polarization \eq{P1gc_def} and the classical zeroth-order guiding-center magnetization \eq{M_0gc}. These results confirm that the two-step perturbation analysis used in modern gyrokinetic theory yield a consistent set of gyrokinetic Vlasov-Maxwell equations that include all first-order polarization and magnetization effects. 

\acknowledgments

This work was supported by a U.~S.~Dept.~of Energy grant under contract No.~DE-FG02-09ER55005.

\appendix

\section{\label{sec:Vlasovia}Reduced Polarization and Magnetization}

In this Appendix, we present a brief summary of the derivation of reduced polarization and magnetization effects \cite{Brizard_Vlasovia} induced by a general near-identity phase-space transformation ${\bf z} \rightarrow \ov{\bf z} \equiv {\cal T}_{\epsilon}{\bf z}$. The dynamical reduction introduced by this phase-space transformation yields the reduced Vlasov equation
\begin{equation}
\pd{\ov{f}}{t} \;+\; \frac{1}{\ov{\cal J}}\;\pd{}{\ov{z}^{\alpha}}\left( \ov{\cal J}\;\frac{d_{\epsilon}\ov{z}^{\alpha}}{dt}\;\ov{f} \right) \;=\; 0,
\label{eq:Vlasov_eq}
\end{equation}
where $\ov{\cal J}$ denotes the Jacobian of the transformation ${\cal T}_{\epsilon}$ and the reduced Hamilton equations $d_{\epsilon}\ov{z}^{\alpha}/
dt \equiv \{ \ov{z}^{\alpha},\; \ov{H}\}_{\epsilon}$ are represented in terms of a reduced Hamiltonian $\ov{H}$ and a reduced Poisson bracket $\{\;,\;
\}_{\epsilon}$ (not necessarily canonical). 

Reduced polarization and magnetization effects are associated with the reduced displacement $\ov{\vb{\rho}}_{\epsilon} \equiv {\sf T}_{\epsilon}^{-1}
{\bf x} \;-\; \ov{{\bf x}}$. We begin with the push-forward derivation of the reduced polarization generated by the phase-space transformation ${\cal T}_{\epsilon}$. The particle charge density (summation over particle species is implied)
\begin{equation}
\varrho \;\equiv\; e\;\int\;\ov{f}\;\delta^{3}(\ov{{\bf x}} + \ov{\vb{\rho}}_{\epsilon} - {\bf r})\;d^{6}\ov{z} \;\equiv\; \ov{\varrho} \;-\; 
\nabla\bdot\ov{{\bf P}}
\label{eq:rho_def}
\end{equation}
is expressed in terms of the reduced charge density
\begin{equation}
\ov{\varrho} \;\equiv\; e\;\int\;\ov{f}\;\delta^{3}(\ov{{\bf x}} - {\bf r})\;d^{6}\ov{z} \;=\; e\;\int\;\ov{f}\;d^{3}\ov{v},
\label{eq:rhobar_def}
\end{equation}
and the reduced polarization charge density $\ov{\varrho}_{\rm pol} \equiv -\,\nabla\bdot\ov{{\bf P}}$, where the reduced polarization is defined as a multipole expansion (dipole + quadrupole + ...) associated with increasing powers of $\ov{\vb{\rho}}_{\epsilon}$:
\begin{equation}
\ov{{\bf P}} \;\equiv\; e\;\int\;\ov{f}\;\ov{\vb{\rho}}_{\epsilon}\;d^{3}\ov{v} \;-\; \nabla\bdot\left( \frac{e}{2}\; \int\;\ov{f}\;
\ov{\vb{\rho}}_{\epsilon}\,\ov{\vb{\rho}}_{\epsilon}\;d^{3}\ov{v} \right) \;+\; \cdots.
\label{eq:pol_def}
\end{equation}

Next, we consider the push-forward derivation of the reduced magnetization generated by the phase-space transformation ${\cal T}_{\epsilon}$. This derivation uses the push-forward transformation of the particle velocity ${\bf v} \equiv d{\bf x}/dt$:
\begin{equation}
\ov{{\bf v}} \;=\; {\sf T}_{\epsilon}^{-1}{\bf v} \;\equiv\; \left[{\sf T}_{\epsilon}^{-1}\left(\frac{d}{dt}\,{\sf T}_{\epsilon}\right)\right]\;
{\sf T}_{\epsilon}^{-1}{\bf x} \;=\; \frac{d_{\epsilon}\ov{{\bf x}}}{dt} \;+\; \frac{d_{\epsilon}\ov{\vb{\rho}}_{\epsilon}}{dt}.
\label{eq:push_v}
\end{equation}
The particle current density
\begin{equation}
{\bf J} \;\equiv\; e\;\int\;\ov{f}\;\left( \frac{d_{\epsilon}\ov{{\bf x}}}{dt} \;+\; \frac{d_{\epsilon}\ov{\vb{\rho}}_{\epsilon}}{dt} \right)\;
\delta^{3}(\ov{{\bf x}} + \ov{\vb{\rho}}_{\epsilon} - {\bf r})\;d^{6}\ov{z} \;=\; \ov{{\bf J}} \;+\; \pd{\ov{{\bf P}}}{t} \;+\; 
c\;\nabla\btimes\ov{{\bf M}},
\label{eq:J-def}
\end{equation}
is thus expressed in terms of the reduced current density
\begin{equation}
\ov{{\bf J}} \;\equiv\; e\;\int\;\ov{f}\;\frac{d_{\epsilon}\ov{{\bf x}}}{dt}\;d^{3}\ov{v},
\label{eq:Jbar_def}
\end{equation}
where $\ov{{\bf J}}_{\rm pol} \equiv \partial\ov{{\bf P}}/\partial t$ is the reduced polarization current density, and the reduced magnetization current density $\ov{{\bf J}}_{\rm mag} \equiv c\,\nabla\btimes\ov{{\bf M}}$. If we expand Eq.~\eq{J-def} in powers of $\ov{\vb{\rho}}_{\epsilon}$, we obtain
\begin{equation}
{\bf J} \;=\; e\;\int\;\ov{f}\;\left( \frac{d_{\epsilon}\ov{{\bf x}}}{dt} \;+\; \frac{d_{\epsilon}\ov{\vb{\rho}}_{\epsilon}}{dt} \right)\;d^{3}\ov{v} \;-\;
\nabla\bdot\left[ e\;\int\;\ov{f}\;\ov{\vb{\rho}}_{\epsilon}\;\left( \frac{d_{\epsilon}\ov{{\bf x}}}{dt} \;+\; 
\frac{d_{\epsilon}\ov{\vb{\rho}}_{\epsilon}}{dt} \right)\;d^{3}\ov{v}\right] \;+\; \cdots.
\label{eq:J_Taylor}
\end{equation}
Next, we take the partial time derivative of the reduced polarization \eq{pol_def}, with the reduced Vlasov \eq{Vlasov_eq}, we can insert
\[ e\;\int\;\ov{f}\;\left( \frac{d_{\epsilon}\ov{\vb{\rho}}_{\epsilon}}{dt} \right)\;d^{3}\ov{v} \;=\; \pd{\ov{\bf P}}{t} \;+\; \nabla\bdot\left\{
e\;\int \ov{f}\;\left[ \frac{d_{\epsilon}\ov{{\bf x}}}{dt}\;\ov{\vb{\rho}}_{\epsilon} \;+\; \frac{1}{2} \left( \ov{\vb{\rho}}_{\epsilon}\;
\frac{d_{\epsilon}\ov{\vb{\rho}}_{\epsilon}}{dt} \;+\; \frac{d_{\epsilon}\ov{\vb{\rho}}_{\epsilon}}{dt}\;\ov{\vb{\rho}}_{\epsilon}\right)\;\right]
\;\right\} \;+\; \cdots \]
into Eq.~\eq{J_Taylor} and obtain
\begin{equation}
{\bf J} \;=\; \ov{\bf J} \;+\; \pd{\ov{\bf P}}{t} \;+\; \nabla\bdot\left\{ e\;\int \ov{f}\;\left[ \left( \frac{d_{\epsilon}\ov{{\bf x}}}{dt}\;
\ov{\vb{\rho}}_{\epsilon} \;-\; \ov{\vb{\rho}}_{\epsilon}\;\frac{d_{\epsilon}\ov{{\bf x}}}{dt} \right) \;+\; \frac{1}{2} \left( 
\frac{d_{\epsilon}\ov{\vb{\rho}}_{\epsilon}}{dt}\;\ov{\vb{\rho}}_{\epsilon} \;-\; \ov{\vb{\rho}}_{\epsilon}\;
\frac{d_{\epsilon}\ov{\vb{\rho}}_{\epsilon}}{dt}\right)\;\right]\;\right\} \;+\; \cdots.
\label{eq:J_pre_mag}
\end{equation}
Lastly, by using the identity $\nabla\bdot({\bf B}{\bf A} - {\bf A}{\bf B}) \equiv \nabla\btimes({\bf A}\btimes{\bf B})$, for two arbitrary vector fields ${\bf A}$ and ${\bf B}$, we obtain an expression for the reduced magnetization
\begin{equation}
\ov{{\bf M}} \;\equiv\; \frac{e}{c}\;\int\;\ov{f}\;\left[\ov{\vb{\rho}}_{\epsilon}\btimes\left( \frac{1}{2}\,
\frac{d_{\epsilon}\ov{\vb{\rho}}_{\epsilon}}{dt} \;+\; \frac{d_{\epsilon}\ov{{\bf x}}}{dt} \right)\right] \;d^{3}\ov{v} \;+\; \cdots.
\label{eq:mag_def}
\end{equation}
In Eq.~\eq{mag_def}, the term associated with $(1/2)\,\ov{\vb{\rho}}_{\epsilon}\btimes d_{\epsilon}\ov{\vb{\rho}}_{\epsilon}/dt$ represents the intrinsic reduced magnetization while the term $\ov{\vb{\rho}}_{\epsilon}\btimes d_{\epsilon}\ov{{\bf x}}/dt$ represents the moving-electric-dipole contribution.

\end{document}